\documentclass{article}
\usepackage{frascatiphys_R,epsf,epsfig}
\begin{document}
\title{WHAT NEXT IN FLAVOUR PHYSICS AND CP VIOLATION?}
\author{John Ellis \\
{\em TH Division, CERN, CH 1211 Geneva 23, Switzerland}\\
CERN-TH/2002-339~~~~~~hep-ph/0211322}
\maketitle
\baselineskip=11.6pt
\begin{abstract}

The future of flavour physics and CP violation in the quark, lepton and
Higgs sectors are discussed, particularly from the viewpoint of physics
beyond the Standard Model, such as supersymmetry. Current issues in $B \to
\pi^+ \pi^-, \phi K_s$ and $D^{*+} D^{*-}$, $B_s$ physics and rare $B$
decays are reviewed. The prospects for seeing flavour and CP violation in
the charged-lepton sector are discussed, using the minimal supersymmetric
seesaw model as a guide. Finally, the possible consequences of CP
violation in the Higgs sector are mentioned.

\end{abstract}
\baselineskip=14pt
\section{Mea Maxima Culpa}
The organizers have asked me to look towards the future, rather than
summarize this meeting. Unfortunately, this is just as well, because
commitments at CERN prevented me from attending most of the meeting. I am
very sorry that I missed many interesting subjects, such as factorization,
$J/\psi$ production at RHIC, charmonium, heavy-quark effective theory,
${\bar b} b$ production, $b$-quark fragmentation, $W \to c {\bar s}$
decay, $B \to \ell \nu \gamma$, Re($\epsilon^\prime / \epsilon$), $K_S \to
\gamma \gamma$, CLEO-c, LHCb light, $Z \to {\bar b} b$, $\Delta G$, $x_s$,
$D \to \sigma, \kappa$, flavour textures, $B \to \ell^+ \ell^{\prime -}$,
$B_s - {\bar B_s}$ mixing, $D_0 - {\bar D_0}$ mixing, $\Delta \Gamma_s
\Gamma_s$, $\tau(D_s) / \tau(D_0)$, $\tau (\Xi^+_c) / \tau (\Lambda^+_c)$,
the $^1$D$_2({\bar b} b)$ and many more .... For these reasons, I could
not in any case present a balanced summary of the meeting.

\section{A Personal Point of View}

There are three preferred experimental arenas for probing flavour dynamics
and CP violation: the quark sector - where both are well established, the
lepton sector - where flavour mixing has been seen among the neutrinos and
CP violation is expected, and the Higgs sector - about which we have no
direct experimental information. Reflecting my personal bias, I assume in
discussing these sectors that supersymmetry will appear at some 
accessible energy.

In the quark sector, dare we hope that that the current triumph of the
standard Kobayashi-Maskawa (KM) model in predicting correctly the value of
$\sin 2 \beta$ observed in $B_0 \to J/\psi K_s$ decays may be short-lived?  
As discussed at this meeting, the first rounds of data on $B \to \pi^+
\pi^-, \phi K_s$ and $D^{*+} D^{*-}$ decay asymmetries do not agree very
well with the KM model. Might one of these be a harbinger of new physics,
such as supersymmetry? Answers to the tough questions are still in the
future: does the unitarity triangle close, or is it a quadrangle? New
tools for analyzing flavour dynamics in the quark sector await us:  what
will $B_s$ physics or $b \to s \gamma, s \ell^+ \ell^-$ tell us?

In the neutrino sector, many questions about neutrino masses and mixing 
remain unanswered: is the large-mixing-angle (LMA) solar solution correct? 
What is the value of $\theta_{13}$? Is there a CP-violating phase 
$\delta$? What are the absolute values of the neutrino masses? Beyond 
neutrinos, in the presence of low-energy supersymmetry we may expect a new 
flavour frontier to open up among the charged leptons: will $\mu \to e 
\gamma, \tau \to e/\mu \gamma, \mu \to 3 e$ and $\tau \to 3 \ell$ be 
observable? Do the electron and muon have measurable CP-violating electric 
dipole moments? What is the relation to leptogenesis?

The final frontier for studies of flavour dynamics and CP violation may be
the Higgs sector, which is their origin in the Standard Model (SM). In the
minimal supersymmetric extension of the Standard Model (MSSM), the masses,
mixings and couplings of the physical Higgs bosons may exhibit observable
flavour- and CP-violating effects.

\section{Roadmap to Physics Beyond the Standard Model}

Let us first set flavour dynamics in the general context of physics beyond
the SM.

The standard list of problems beyond the SM includes those of {\it
Unification} - can one find a single simple framework for all the gauge
interactions?  {\it Flavour} - why so many different types of quarks and
leptons and what explains their patterns of mixing and CP violation? and
{\it Mass} - do particle masses really originate from a Higgs boson, and
if so why are they so small, where there may be a r\^ole for
supersymmetry? Beyond all these `beyonds' there is the quest for a {\it
Theory of Everything}, capable of reconciling gravity with quantum
mechanics as well as solving all the above problems, perhaps via
superstring or M theory?

At what energy scales might appear these examples of new physics? LEP told
us that they cannot appear below 100~GeV, and quantum gravity must become
strong by $10^{19}$~GeV at the latest. Within this range, we believe that
the problem of mass must be resolved at some energy below about 1~TeV, by
the discovery of a Higgs boson and/or supersymmetry. Measurements of gauge
couplings give circumstantial support to supersymmetric grand unification
at around $10^{16}$~GeV with sparticles appearing around 1~TeV.  {\it
However, we have little, if any, idea of the scale at which the flavour
problem may be solved}. Perhaps only at the quantum-gravity scale $\sim
10^{19}$~GeV?  perhaps at the GUT scale $\sim 10^{16}$~GeV? perhaps at
some intermediate scale, as suggested by the seesaw model of neutrino
masses? perhaps at the TeV scale? How far along the road will we solve 
flavour dynamics and the find the origin of CP violation?

\section{Milestones in CP Violation}

\begin{figure}
\hspace{1cm}   
\epsfig{figure=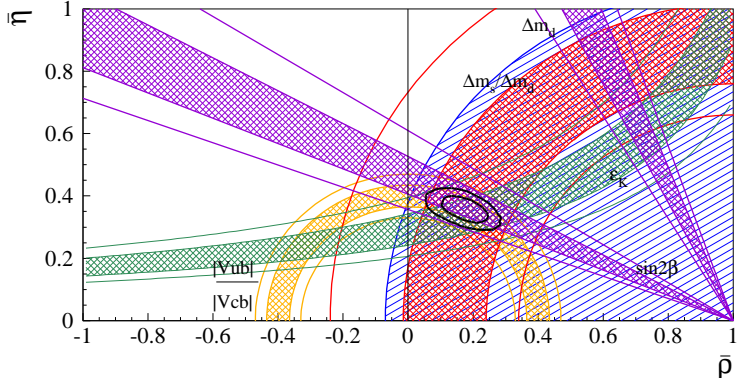,width=10cm}
\hglue3.5cm
\caption[]{\it 
A global fit to the unitarity triangle~\cite{Stocchi}, demonstrating good 
agreement with the measurements of $\sin 2 \beta$ by BaBar and 
Belle~\cite{twobeta}.}
\label{fig:CKM}
\end{figure}

Our progress along this road can be measured by a plethora of milestones.
Long after its discovery in the $K^0$ mass matrix via $K^0 \to \pi^+
\pi^-$ decay, we have only recently passed two important ones:

$\bullet$ The measurement of direct CP violation in $K^0 \to 2 \pi$ decay 
amplitudes~\cite{epsprime}, as long predicted in the KM model~\cite{EGN},

$\bullet$ Observation of CP violation elsewhere, namely in $B^0 \to J/\psi 
K_s$ decay~\cite{twobeta}.\\
The latest NA48 and KTeV measurements of Re($\epsilon^\prime / \epsilon$) 
are now in relatively good agreement: $(14.7 \pm 2.2) \times 10^{-4}$ and 
$(20.7 \pm 2.8) \times 10^{-4}$, leading to the world 
average~\cite{epsprime}:
\begin{equation}
{\rm Re} (\epsilon^\prime / \epsilon) \; = \; (16.6 \pm 1.6 ... 2.3) 
\times 10^{-4},
\label{epsprime}
\end{equation}
where the first error is naive, and the second one is rescaled according 
to the Particle Data Group prescription. The value 
(\ref{epsprime}) is 
consistent with theoretical calculations, but these are not very accurate, 
because of delicate cancellations between different non-perturbative 
matrix elements. Measurements of $\sin 2 \beta$ are already startlingly 
precise~\cite{twobeta}:
\begin{equation}
\sin 2 \beta \; = \; 0.741 \pm 0.067,
\label{sin2beta}
\end{equation}
and very consistent with KM expectations of mixing-induced CP violation, 
as seen in Fig.~\ref{fig:CKM}~\cite{Stocchi}. 
However, the origin of the CP asymmetry in 
$B^0 \to J/\psi K_s$ decay is not yet confirmed, hence the importance of 
the next milestone, namely:

$\bullet$ The measurement of direct CP violation in $B^0 \to 2 \pi$ decay  
amplitudes, predicted to be the angle $\alpha ( = \phi_2)$ in the KM 
model.\\
As discussed later in more detail, the search for this effect~\cite{pipi} 
is currently 
the subject of some discussion~\cite{Browder}. Beyond it, many other 
CP-violating milestones 
beckon:

$\bullet$ CP violation in other $K$ decays, such as $K^0_L \to \pi^0 {\bar 
\nu} \nu$ decay,

$\bullet$ CP violation in other $B$ decays, such as the measurement of the 
third unitarity angle $\gamma$,

$\bullet$ CP violation in $D$ decays. \\
As we heard at this meeting, there is no hint of CP violation in $D^0 - 
{\bar D^0}$ mixing~\cite{noDCPV}, which is expected only at a very low 
level in the SM, 
making it an excellent place to look for new physics beyond it. Other 
places to look for new sources of CP violation include

$\bullet$ The neutron electric dipole moment $d_n$,

$\bullet$ CP violation in neutrino oscillations via the MNS phase 
$\delta$,

$\bullet$ T violation in lepton decays such as $\mu \to 3 e$ and $\tau \to 
3 \ell$,

$\bullet$ The lepton electric dipole moments $d_e, d_\mu, d_\tau$. \\
Only after we pass some more of these milestones will we have a chance of 
pinning down the origin(s) of CP violation: is it due to the KM mechanism 
alone? or are there other contributions? perhaps due to $\theta_{QCD}$? 
the MNS phase? supersymmetry? or ...?

\section{The Next Steps along the CP Road}

\subsection{Quo Vadis $B^0 \to \pi^+ \pi^-$?}

\begin{figure}
\hspace{1.5cm}
\epsfig{figure=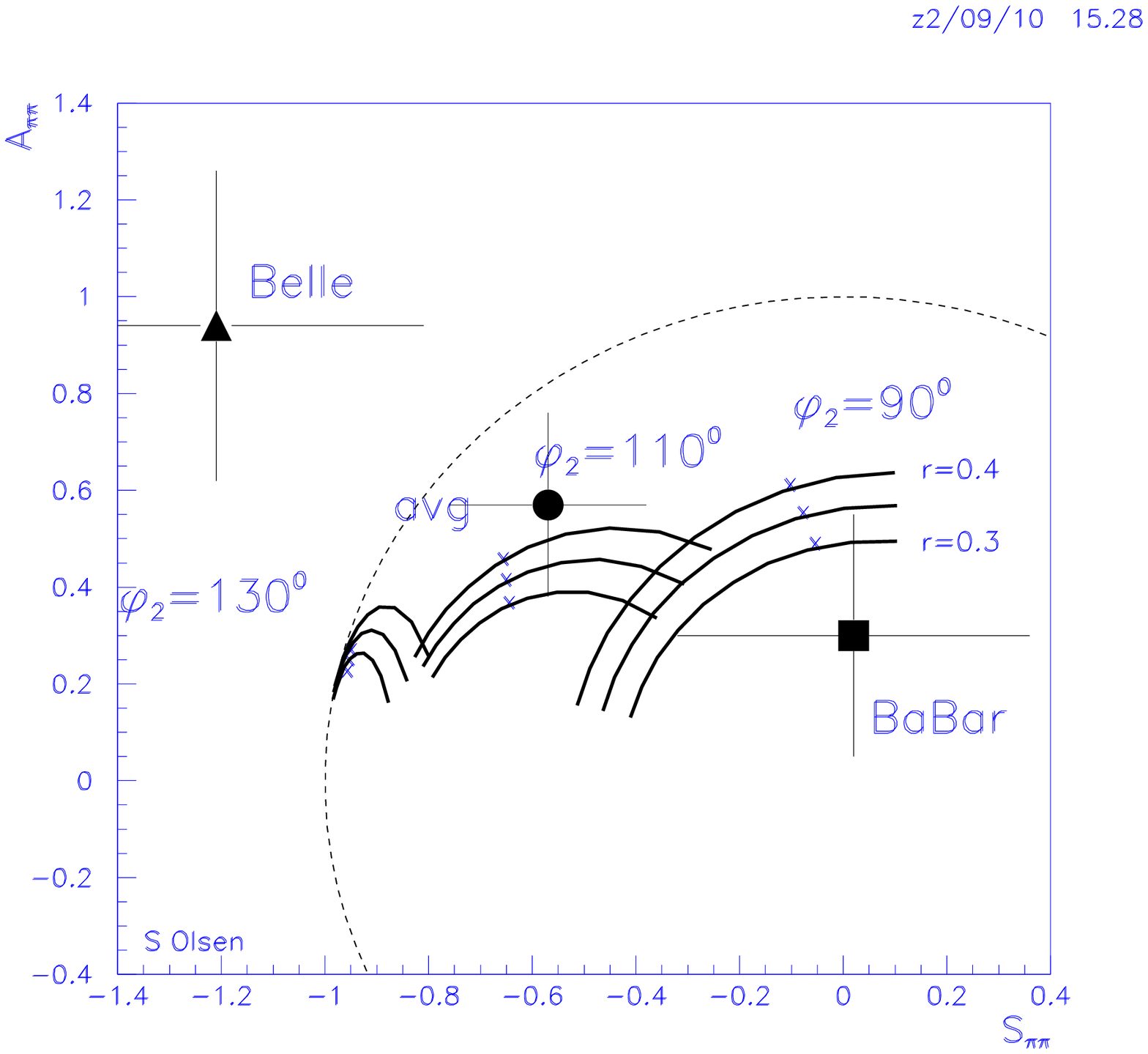,width=7cm}
\hglue3.5cm
\caption[]{\it 
The BaBar and Belle measurements of the asymmetry parameters 
$S_{\pi\pi},A_{\pi \pi}$~\cite{pipi}, and their naive average, are 
compared with KM 
predictions for 
different values of $\phi_2 = \alpha$, the penguin pollution factor $r$ 
and the strong-interaction phase. Also shown is the unitarity limit 
$A_{\pi \pi}^2 + S_{\pi \pi}^2 = 1$~\cite{Browder}.} 
\label{fig:pipi} 
\end{figure}

As you know, this decay mode receives contributions from $b \to u {\bar
u} d$ tree diagrams and $b \to s {\bar u} s$ penguin diagrams, which contain 
both a weak and a strong phase. The resulting CP-violating asymmetry contains 
two parts:
\begin{equation}
S_{\pi \pi} \sin ( \Delta m_d \Delta t ) \; + \; 
A_{\pi \pi} \cos ( \Delta m_d \Delta t ),
\label{B2pi}
\end{equation}
where the latter term is that due to direct CP violation. The values of 
$S_{\pi \pi}$ and $A_{\pi \pi}$ depend on the proportion of penguin 
pollution $r$ (that may be constrained by other measurements such as 
$B^0 \to 2 \pi^0$ and $B^+ \to K_s \pi^+$) and as well the angle $\alpha$ 
(or $\phi_2$) that we seek to determine. As seen in 
Fig.~\ref{fig:pipi}~\cite{Browder}, 
the first measurements by BaBar and Belle~\cite{pipi} are not in good 
agreement, 
though the naive average suggests that $S_{\pi \pi} \sim - 0.6, A_{\pi 
\pi} \sim 0.6$, which are consistent with $\phi_2$ (or $\alpha$) $\sim 
110$ degrees, as also seen in Fig.~\ref{fig:pipi}. Naive 
averaging may not be adequate, however, since the 
Belle measurement lies outside the physical boundary: $A_{\pi \pi}^2 + 
S_{\pi \pi}^2 = 1$, which should be taken into account in any fit.

There has recently been much progress in calculating exclusive $B$ decay
amplitudes using the QCD factorization framework~\cite{BBNS}, with error 
estimates based on evaluations of power corrections and annihilation 
diagrams. This framework suggests $S_{\pi \pi} \sim 
-0.3$ to -0.9~\cite{BBNS}, consistent with the naive average shown in 
Fig.~\ref{fig:pipi}. Thus there is reason to hope that $B^0 \to \pi^+ 
\pi^-$ decay could become a valuable check on the KM model, as soon as the 
experimental situation settles down.

\subsection{Quo Vadis $B \to \phi K_s$?}

In the KM model, this decay is mediated by a strange gluonic penguin
diagram: $b \to s + (g \to {\bar s} s)$, which has no intrinsic weak
phase. Therefore, this decay should exhibit only mixing-induced CP
violation, and should have the same asymmetry $\sin 2 \beta$ as $B \to
J/\psi K_s$. Other processes mediated by the same diagram include $B \to
(\eta^\prime, K^+ K^-) K_s$~\cite{Sokoloff}, and first 
measurements 
of these decay
asymmetries are consistent (within large errors) with that in $B \to 
J/\psi K_s$ and the KM model:
\begin{equation}
\eta^\prime: \; 0.76 \pm 0.36; \; K^+ K^-: \; 0.52 \pm 
0.47,
\label{etaKK}
\end{equation}
whereas the decay asymmetry in  $B \to \phi K_s$ looks rather 
different:
\begin{equation}
\phi: \; - 0.39 \pm 0.41.
\label{phi}
\end{equation}
If this result holds up with more statistics, it would require new physics 
in the $b \to s + (g \to {\bar s} s)$ penguin diagram.

Several theoretical papers have appeared since the result (\ref{phi})  
emerged, discussing models based on conventional $R$-conserving
supersymmetry~\cite{Silvestrini}, $R$-violating supersymmetry~\cite{who},
left-right symmetry~\cite{Raidal} and a $Z^\prime$ model~\cite{Hiller}.
{\it Une affaire \`a suivre ....}

\subsection{Quo Vadis $B \to D^{*+} D^{*-}$?}

The dominant diagram contributing to this process is thought to be $b \to 
c + (W \to {\bar c} d)$, with the competing penguin diagram $b \to d + (g 
\to {\bar c} c)$ thought to be rather small: $|P / T| < 0.1$. A first 
measurement of the CP-violating asymmetry -Im($\lambda_+$) that should 
conicide with $\sin 2 \beta$ yields~\cite{DD}:
\begin{equation}
- {\rm Im} (\lambda_+) \; = \; - 0.31 \pm 0.43 \pm 0.1,
\label{Dstar}
\end{equation}
which deviates by about 2.7 $\sigma$, nominally. However, the 
experimentalists caution that, with the current low statistics, the errors 
are not Gaussian. {\it Une autre affaire \`a suivre ....}

\subsection{Quo Vadis $\gamma$?}

There are various isospin relations between $B \to \pi K$ amplitudes that
can be used to provide information about $\gamma$: e.g., the relation
between those for the charged $B^+ \to \pi^0 K^+, \pi^+ K^0$, the relation
between those for the neutral $B^0 \to \pi^- K^+, \pi^0 K^0$, and the
mixed relation between $B^0 \to \pi^- K^+$ and $B \to \pi^+ K^0$. The 
charged amplitudes may be parametrized by the two 
quantities~\cite{Fleischer}
\begin{equation}
R^c, A_0^c \; \equiv \; 2 {B(\pi^0 K^+) \pm B(\pi^0 K^-) \over 
B(\pi^+ K^0) + B(\pi^- {\bar K^0})},
\label{RcA0c}
\end{equation}
which depend on the strong tree-to-penguin ratio $r_c (\sim 0.2?)$, 
the electroweak tree-to-penguin ratio $q (\sim 0.7?)$, and the difference 
$\delta_c$ between the tree- and penguin-diagram phases.

\begin{figure}[t]
\vspace*{-0.1cm}
$$\hspace*{-1.cm}
\epsfysize=0.16\textheight
\epsfxsize=0.26\textheight
\epsffile{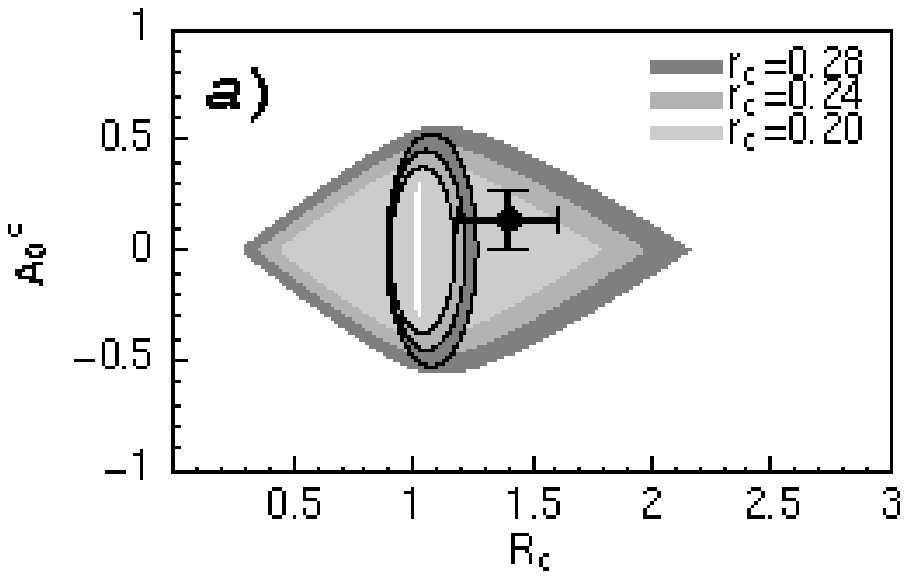} \hspace*{0.3cm}
\epsfysize=0.16\textheight
\epsfxsize=0.26\textheight
\epsffile{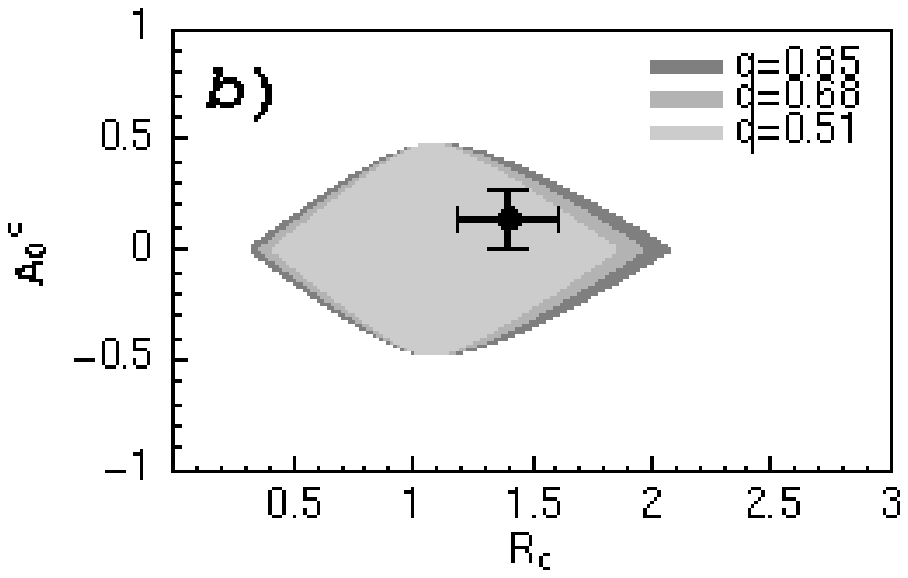}
$$
\vspace*{-0.3cm}
$$\hspace*{-1.cm}
\epsfysize=0.16\textheight
\epsfxsize=0.26\textheight
\epsffile{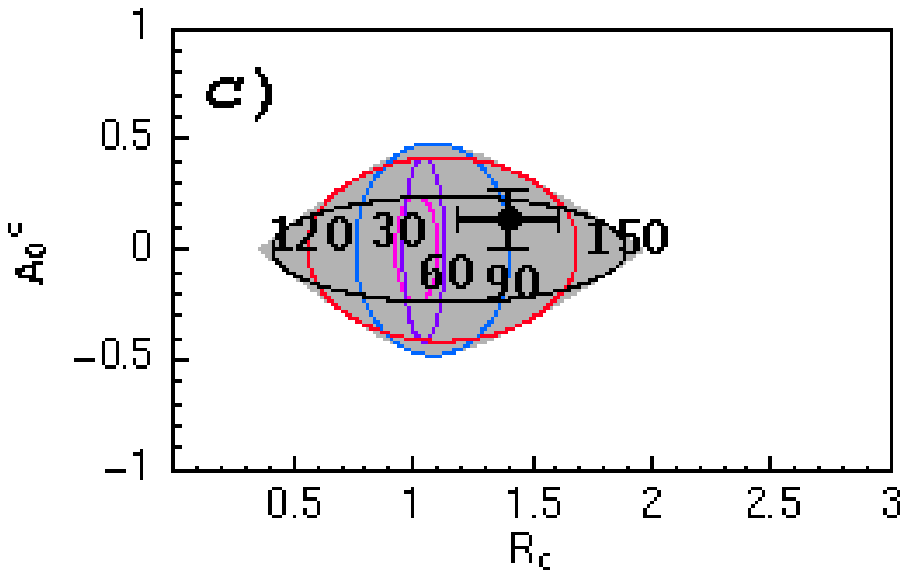} \hspace*{0.3cm}
\epsfysize=0.16\textheight
\epsfxsize=0.26\textheight
\epsffile{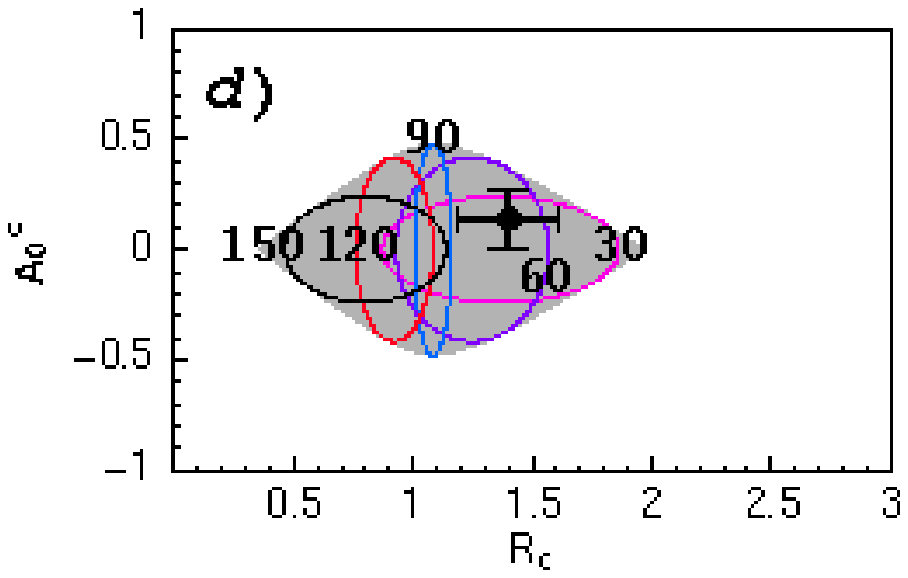}
$$
\caption[]{\it Allowed regions in the $R_{\rm c}$--$A_0^{\rm c}$ 
plane for charged $B \to \pi K$ decays, showing the effects of varying (a)
the strong penguin pollution factor $r_c$, (b) the electroweak penguin 
pollution factor $q$, (c) the KM phase $\gamma$ and (d) the phase difference 
$\delta_c$~\cite{Fleischer}.}
\label{fig:Fleischer}
\end{figure}

Fig.~\ref{fig:Fleischer} shows the current status of measurements of $R^c$
and $A_0^c$~\cite{Fleischer}. We see from the third panel that the data
prefer $\gamma > 90$ degrees, whereas the global KM fit shown in
Fig.~\ref{fig:CKM} prefers $\gamma < 90$ degrees. Again, it remains to see
whether this possible discrepancy is confirmed by more data on the same
decay modes, and/or on other decays such as $B^- \to D^0 K^-$, $B^0 \to
D^{(*)\pm} \pi^\pm$, etc.

\subsection{The Road Ahead for $B$ Factories?}

Measurements of $\beta$ at $B$ factories are likely to attain an accuracy 
of $\pm 1$ degree, those of $\alpha$ may reach $\pm 5$ degrees, and those 
of $\gamma$ may reach $\pm 25$ degrees, which would correspond to a check 
of the unitarity triangle at the 15~\% level. It would clearly be 
desirable to push the experimental statistical errors down until they 
match the theoretical systematic errors. This provides worthwhile 
objectives for the subsequent generation of LHCb, BTeV and super-$B$ 
factory experiments.

\section{The $B_s$ Road to CP Violation}

\begin{figure}
\hspace{2cm}   
\epsfig{figure=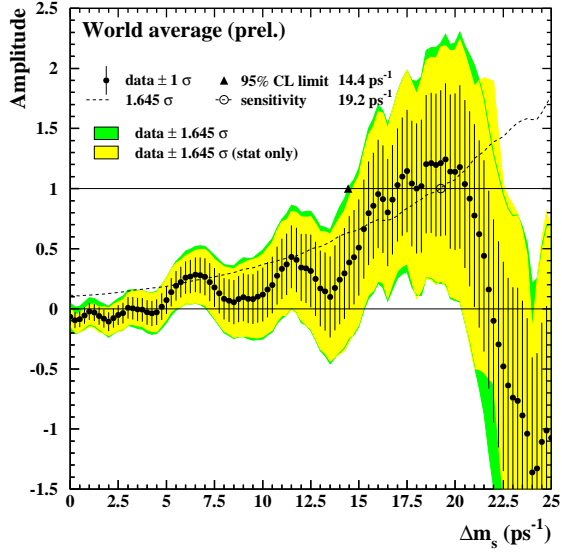,width=8cm}
\hglue3.5cm
\caption[]{\it Combined LEP/SLD/CDF results for $\Delta {m}_s$.
The data are shown as points with error bars, the lines show 
the 95\% C.L. curves, and the dotted curve shows the expected 
sensitivity~\cite{Bsmix}.}
\label{fig:Bsmix}
\end{figure}

There are just three neutral-meson systems where one can reasonably expect 
to see mixing and CP-violating effects in the SM and its plausible 
extensions: the $K^0 - {\bar K^0}$ system that has been explored for many 
decades, the $B^0 - {\bar B^0}$ system that is now being explored at $B$ 
factories, and the the $B_s^0 - {\bar B_s^0}$ system. If we really do 
understand the SM and CP violation as well as we think, we can make many 
reliable predictions for the $B_s^0 - {\bar B_s^0}$ system. Conversely, 
this may be a valuable laboratory for testing the SM, since any deviation 
from these confident predictions would be good evidence for physics beyond 
the SM. Among these predictions, one may list~\cite{Fleischer}:

$\bullet$ A large mixing parameter $x_s \equiv \Delta m_s / \Gamma_s =
{\cal O}(20)$ - this prediction may be on the verge of being confirmed, as
the compilation of present experiments on $B_s^0 - {\bar B_s^0}$ mixing
shown in Fig.~\ref{fig:Bsmix} shows quite a hint of mixing with
approximately the predicted value of $x_s$~\cite{Bsmix};

$\bullet$ The $B_s^0 - {\bar B_s^0}$ mixing phase should be very small: 
$\phi_s = {\rm Arg}(V_{ts}^* V_{tb}) \sim - 2$ degrees;

$\bullet$ There may be a sizeable difference in the total decay widths of
the mass eigenstates of the $B_s^0 - {\bar B_s^0}$ system: $\Delta
\Gamma_s / \Gamma_s \sim 10$~\%.\\
Among the interesting $B_s^0$ decay modes, let us mention $B_s^0 \to 
J/\psi \phi$, whose CP-violating asymmetry should be
\begin{equation}
A_{CP} \simeq \sin (\Delta m_s t) \sin \phi_s,
\label{Jpsiphi}
\end{equation}
and hence very small in the SM. This makes it a good place for new 
contributions to $B_s^0 - {\bar B_s^0}$ mixing, as might occur in 
supersymmetry, for example. Another interesting decay mode is $B_s^0 \to 
D_s^\pm K^\mp$, whose CP-violating asymmetry should be proportional to 
$\phi_s + \gamma$, and hence could (within the SM) be a good way to 
measure $\gamma$.

There are currently no plans to try to accumulate large samples of $B_s^0$
mesons at the operating $B$ factories, so $B_s^0$ physics may be left as
the hunting preserve of the hadronic experiments LHCb~\cite{LHCb} and
BTeV~\cite{BTeV}.

\section{The Supersymmetric Flavour and CP Problems}

In the supersymmetric limit, flavour mixing in the MSSM is identical to
that in the SM, but supersymmetry must be broken. It is commonly thought
that this occurs via gaugino masses $M_a$, scalar mass-squared parameters
$(m_0^2)^i_j$ and trilinear couplings $A_{ijk}$. The gaugino mass 
parameters might have CP-violating phases that could show up in electric 
dipole moments and/or the Higgs sector, as discussed later. The big 
questions concerning $(m_0^2)^i_j$ and $A_{ijk}$ are whether they are 
universal, or at least can be diagonalized in the same basis as the quark 
and lepton flavours, and whether they contain extra CP-violating phases. 
Is the super-CKM mixing of squarks the same as the KM mixing of quarks? If 
not, how does it differ, and why?

Three generic classes of options can be distinguished~\cite{Masiero}:

$\bullet$ Minimal flavour violation, in which the $(m_0^2)^i_j$ and
$A_{ijk}$ {\it are} universal at the GUT scale, being renormalized at
lower energies by the Yukawa couplings $\lambda_{ijk}$, and resulting in a
super-CKM mixing pattern that is related to, and derivable from, the
conventional CKM mixing;

$\bullet$ Extra supersymmetric loop effects, that may in general be
parameterized as quark mass insertions
$(\delta^{d,u}_{ij})_{LL,RR}$~\cite{Silvestrini};

$\bullet$ Extra tree-level effects, as could arise from generic 
$R$-violating interactions.

\begin{figure}
\hspace{2cm}   
\epsfig{figure=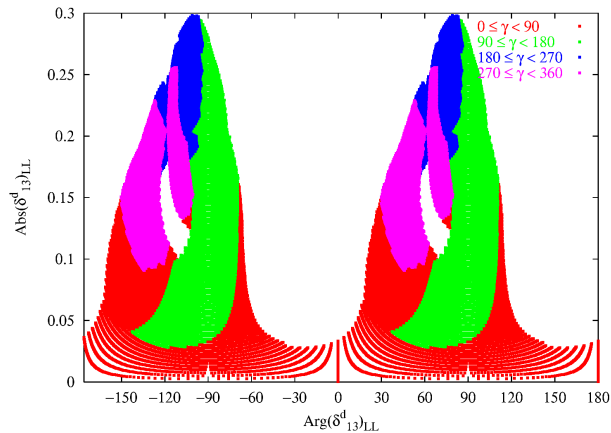,width=8cm}
\hglue3.5cm
\caption[]{\it Phenomenological bounds on the magnitude and phase of the 
insertion $(\delta^d_{13})_{LL}$~\cite{Silvestrini}.}
\label{fig:bounddeltas}
\end{figure}

Quite frankly, fundamental theory provides no clear guidance which option 
Nature might have chosen. On the other hand, the observed suppressions of 
flavour-changing neutral interactions put severe constraints on 
$R$-violating models, which will not be discussed further here. These 
constraints certainly favour models with minimal flavour violation, 
although the best one can do phenomenologically, in a model-independent way, 
is to set upper bounds on the insertions $(\delta^{d,u}_{ij})_{LL,RR}$, as 
exemplified in Fig.~\ref{fig:bounddeltas}~\cite{Silvestrini}.

If supersymmetric flavour violation is indeed minimal, one expects the 
squarks to be approximately degenerate, apart from the $\tilde t$ and 
possibly the $\tilde b$. These loopholes open up interesting opportunities 
in $B$ physics. For example, there could be significant supersymmetric 
contributions to the mass differences $\Delta m_d$ and $\Delta m_s$, 
though not to the ratio $\Delta m_d / \Delta m_s$. These would generate 
knock-on effects in the global unitarity triangle fits and $B_s$ physics. 
Rare $B$ decays already provide interesting upper limits on supersymmetric 
flavour violation and opportunities for the future, as we discuss next.

\section{Rare $B$ Decays}

This is a very rich area~\cite{Ali}, and just a few examples are given 
here.

\begin{figure}[htb]
\epsfig{figure=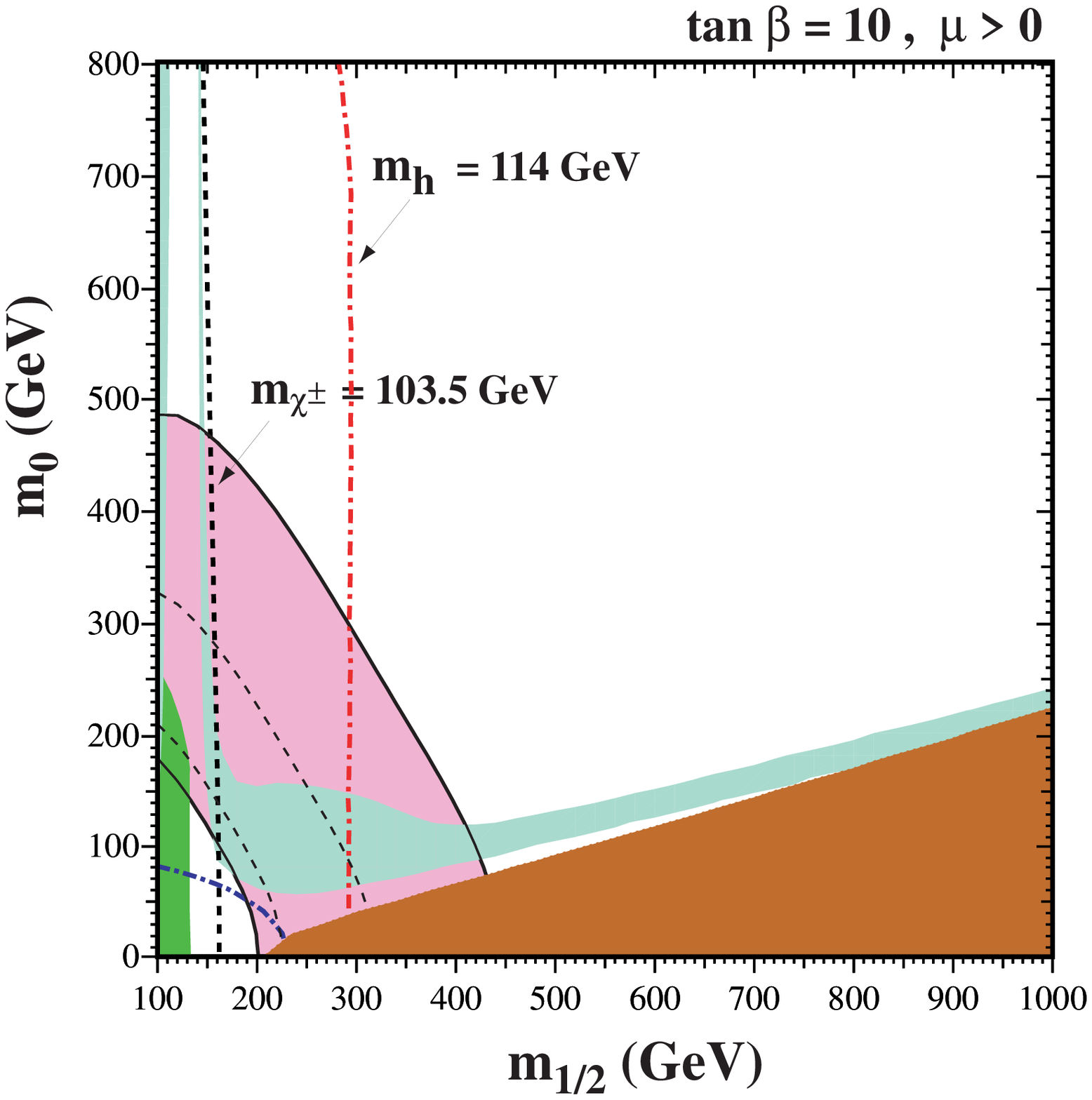,width=6cm}
\epsfig{figure=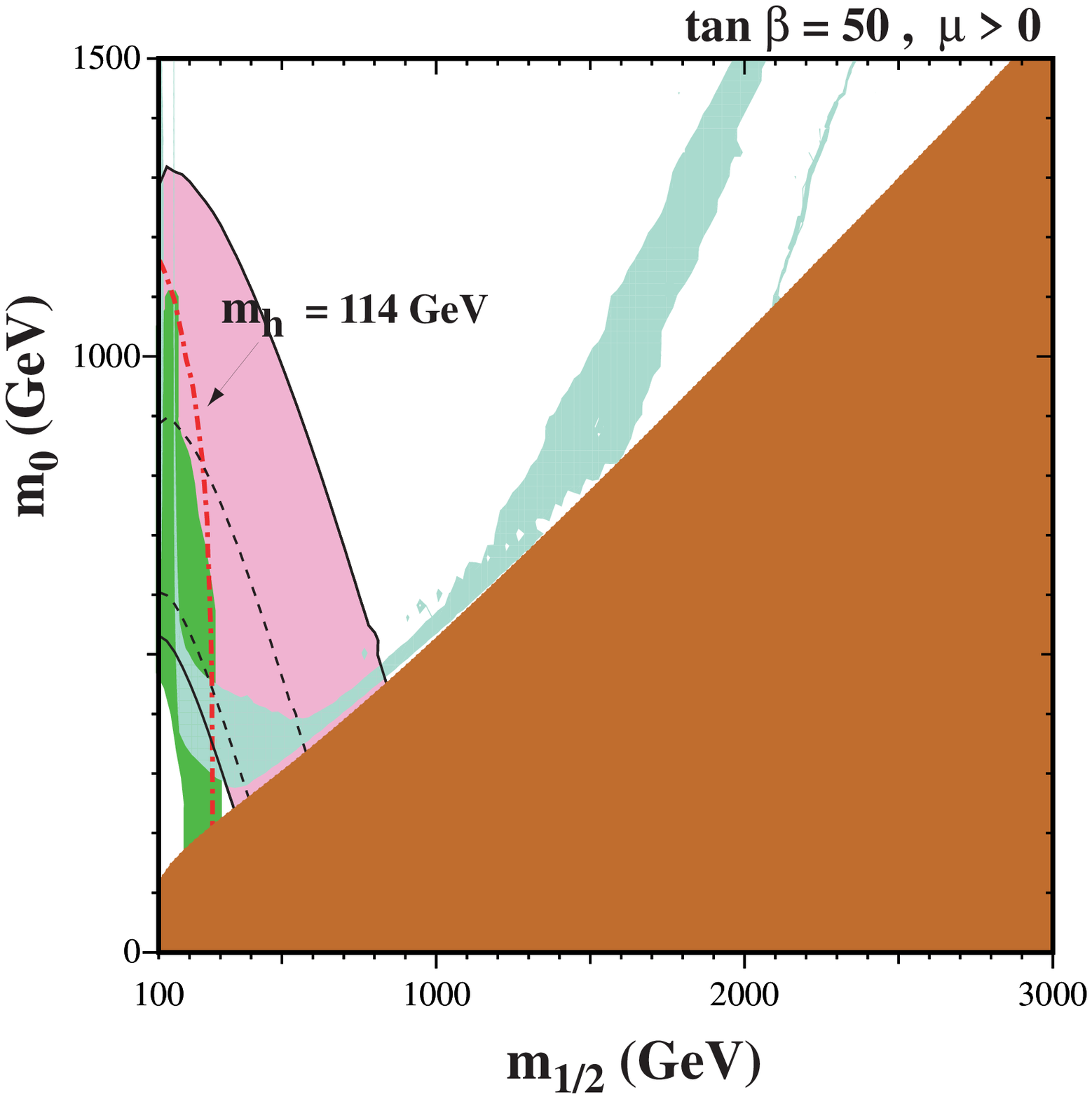,width=6cm}
\caption[]{\it
Compilations of phenomenological constraints on supersymmetry for (a)
$\tan \beta = 10, \mu > 0$, (b) $\tan \beta = 50, \mu > 0$~\cite{EFOS}. 
The
near-vertical lines are the LEP limits $m_{\chi^\pm} = 103.5$~GeV (dashed
and black), shown in (a) only, and $m_h = 114$~GeV (dotted
and red). Also, in the lower left corner of (a), we show the
$m_{\tilde e} = 99$ GeV contour. In the dark
(brick red) shaded regions, the LSP is the charged ${\tilde \tau}_1$, so
this region is excluded. The light (turquoise) shaded areas are the
cosmologically preferred regions with \protect\mbox{$0.1\leq\Omega h^2\leq
0.3$}. The shaded (pink) regions are the $\pm 2 \, \sigma$
ranges of $g_\mu - 2$.
}
\label{fig:CMSSM}
\end{figure}

$\bullet$ $b \to s \gamma$ decay: This process may receive significant 
contributions from the exchanges of charged Higgs bosons $H^\pm$ and 
chargino spartners of the $W^\pm$ and $H^\pm$~\cite{bsg}, and the fact the 
observed 
decay rate agrees within errors with the SM provides important constraints 
on the MSSM parameters, as seen in Fig.~\ref{fig:CMSSM}~\cite{EFOS}. 
In time, one 
could hope to measure a CP-violating asymmetry in $b \to s \gamma$ decays, 
and verify whether $\sin 2 \beta$ measured at the loop level coincides 
with the value measured in the $J/\psi K_s$ decay mode.

\begin{figure}
\hspace{1cm}   
\epsfig{figure=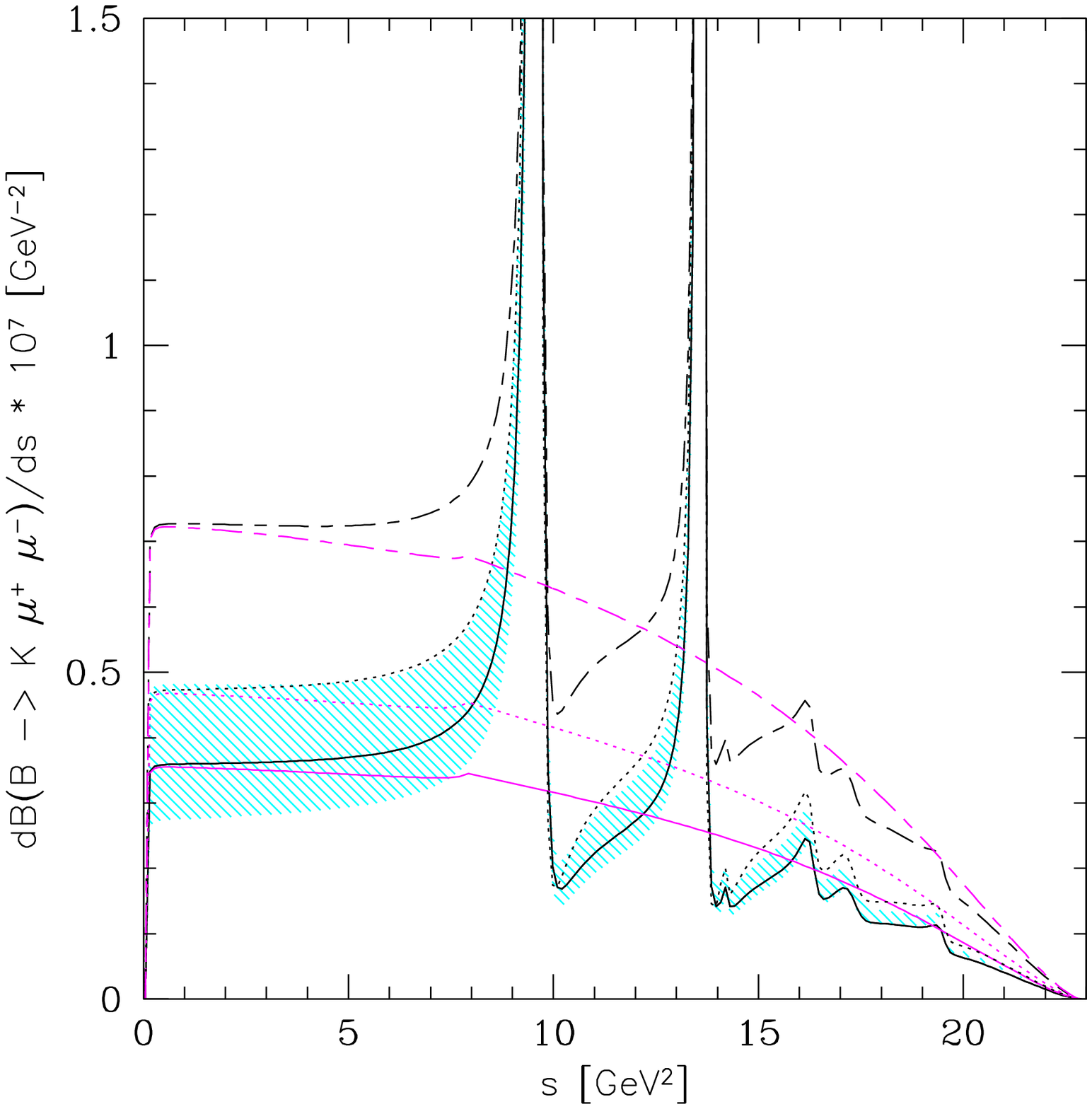,width=5cm}
\epsfig{figure=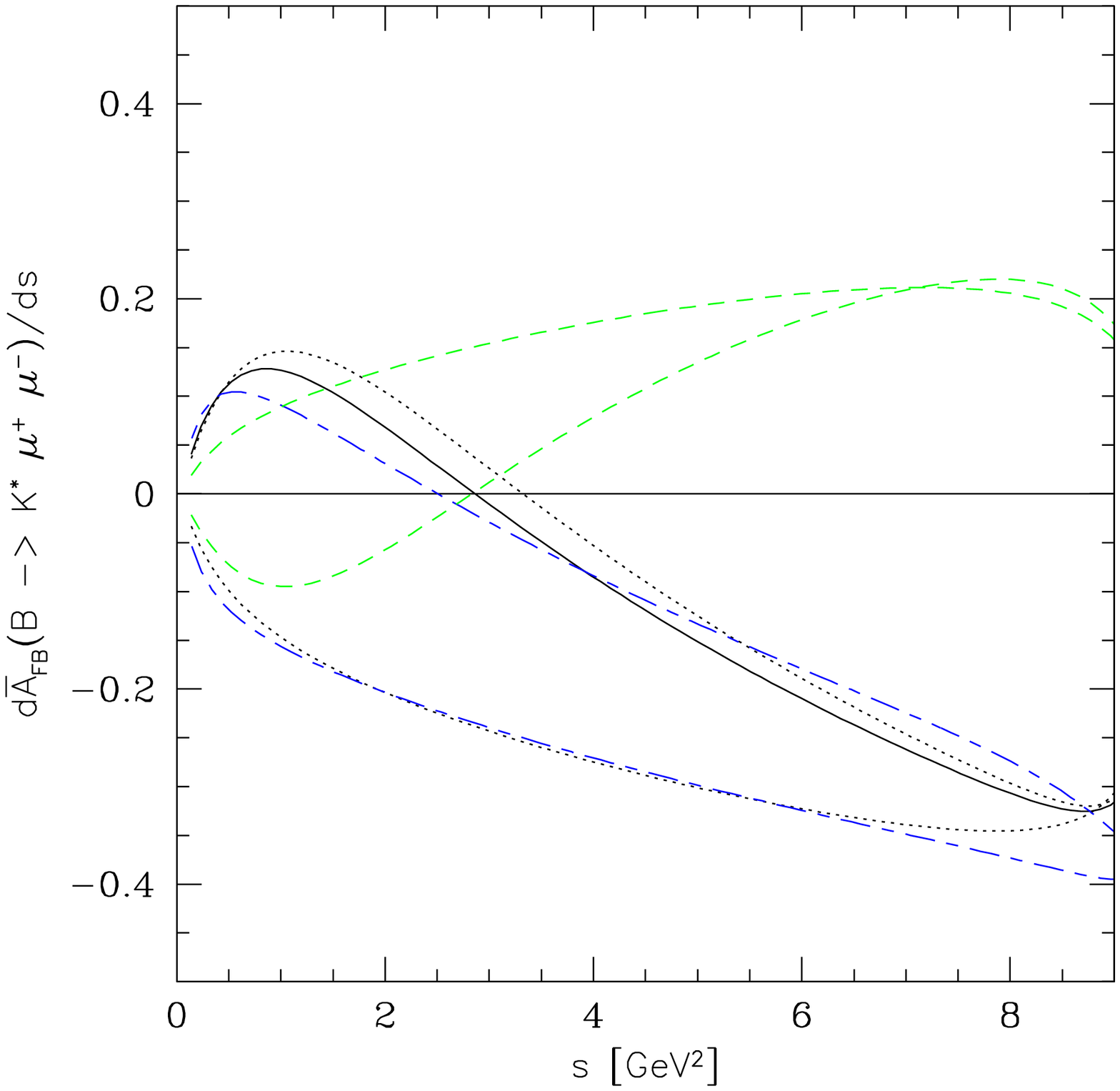,width=5cm}
\hglue3.5cm
\caption[]{\it Possible supersymmetric effects on (a) the invariant-mass 
distribution in $B \to K \mu^+ \mu^-$ decay and (b) the forward-backward 
asymmetry in $B \to K^* \mu^+ \mu^-$ decay~\cite{AL}.}
\label{fig:mellell}
\end{figure}

$\bullet$ $b \to s \ell^+ \ell^-$ decay: The case where the $s$ quark 
yields a $K$ meson has been observed, but not where it yields an excited state 
$K^*$. There could in principle be supersymmetric effects on the total 
decay rate, on the $\ell^+ \ell^-$ mass spectrum, as seen in 
Fig.~\ref{fig:mellell}(a), and on the forward-backward asymmetry $A_{FB}$, 
as seen in Fig~\ref{fig:mellell}(b)~\cite{AL}. 
Again, the question whether $\sin 2 
\beta$(loop) concides with $\sin 2 \beta$($J/\psi K_s$) can be posed.

$\bullet$ $b \to s {\bar \nu} \nu$ decay: This process can also be 
calculated reliably in the SM, and observation of $B \to X_s +$ nothing 
would be interesting for constraining extensions of the SM.

$\bullet$ $B \to \mu^+ \mu^-$ decay: This can receive important 
supersymmetric corrections~\cite{Dreiner}, in particular for larger values 
of $\tan \beta$.

$\bullet$ $B \to \tau^+ \tau^-$ decay: This offers some prospects for 
studying CP violation in the MSSM~\cite{DP}.

$\bullet$ $B \to \tau^\pm \mu^\mp, e^\pm e^\mp$ and $\mu^\pm e^\mp$ 
decays: These could in principle provide interesting windows on flavour 
violation in the lepton sector~\cite{DER}, which is the subject of the next 
section.

It may be interesting to note some of the statistics that may be provided 
by present and forthcoming experiments. For $B \to K^* \gamma$, we may 
expect 6,000 events at the $B$ factories, 25,000 at LHCb or BTeV, and 
120,000 at a super-$B$ factory. The corresponding numbers for $B \to X_s 
\mu^+ \mu^-$ are 120, 4,500 and 6,000, respectively, whilst for $B \to X_s 
{\bar \nu} \nu$ they are 8, 0 and 160, respectively. There is ample 
justification for another generation of $B$ experiments even beyond LHCb 
and BTeV.

\section{Neutrino Flavour Violation}

There is no good reason why either the total lepton number $L$ or the
individual lepton flavours $L_{e, \mu, \tau}$ should be
conserved. We have learnt that the only conserved quantum
numbers are those associated with exact gauge symmetries, just as the
conservation of electromagnetic charge is associated with $U(1)$ gauge
invariance. On the other hand, there is no exact gauge symmetry associated
with any of the lepton numbers.

Moreover, neutrinos have been seen to oscillate between their different
flavours~\cite{SK,SNO}, showing that the separate lepton flavours $L_{e,
\mu, \tau}$ are indeed not conserved, though the conservation of total
lepton number $L$ is still an open question. The observation of such
oscillations strongly suggests that the neutrinos have different masses.
Again, massless particles are generally associated with exact gauge
symmetries, e.g., the photon with the $U(1)$ symmetry of the Standard
Model, and the gluons with its $SU(3)$ symmetry. In the absence of any
leptonic gauge symmetry, non-zero lepton masses are to be expected, in
general.

The conservation of lepton number is an accidental symmetry of the
renormalizable terms in the Standard Model lagrangian. However, one could
easily add to the Standard Model non-renormalizable terms that would  
generate neutrino masses, even without introducing a `right-handed'
neutrino field. For example, a non-renormalizable term of the
form~\cite{BEG}
\begin{equation}
{1 \over M} \nu H \cdot \nu H,
\label{nonren}
\end{equation}
where $M$ is some large mass beyond the scale of the Standard Model, would
generate a neutrino mass term:
\begin{equation}
m_\nu \nu \cdot \nu:
\; m_\nu \; = \; {\langle 0 \vert H \vert 0 \rangle^2 \over M}.
\label{smallm}
\end{equation}
Of course, a non-renormalizable interaction such as (\ref{nonren}) seems
unlikely to be fundamental, and one should like to understand the origin 
of the large mass scale $M$.

The minimal renormalizable model of neutrino masses requires the
introduction of weak-singlet `right-handed' neutrinos $N$. These will in
general couple to the conventional weak-doublet left-handed neutrinos via
Yukawa couplings $Y_\nu$ that yield Dirac masses $m_D \sim m_W$. In
addition, these `right-handed' neutrinos $N$ can couple to themselves via
Majorana masses $M$ that may be $\gg m_W$, since they do not require
electroweak summetry breaking. Combining the two types of
mass term, one obtains the seesaw mass matrix~\cite{seesaw}:
\begin{eqnarray}
\left( \nu_L, N\right) \left(
\begin{array}{cc}
0 & M_{D}\\
M_{D}^{T} & M
\end{array}
\right)
\left(
\begin{array}{c}
\nu_L \\
N
\end{array}
\right),
\label{seesaw}
\end{eqnarray}
where each of the entries should be understood as a matrix in generation
space.

This seesaw model can accommodate the neutrino mixing seen experimentally,
and naturally explains the small differences in the masses-squared of the
light neutrinos. By itself, it would lead to unobservably small
transitions between the different charged-lepton flavours. However,
supersymmetry may enhance greatly the rates for processes violating the
different charged-lepton flavours, rendering them potentially observable,
as we discuss below.

The effective mass matrix for light neutrinos in the seesaw model may be
written as:   
\begin{equation}
{\cal M}_\nu \; = \; Y_\nu^T {1 \over M} Y_\nu v^2 \left[ \sin^2 \beta  
\right]
\label{seesawmass}
\end{equation}
where we have used the relation $m_D = Y_\nu v \left[ \sin \beta \right]$
with $v \equiv \langle 0 \vert H \vert 0 \rangle$, and the factors of  
$\sin \beta$ appear in the supersymmetric version of the seesaw model. 
Diagonalizing the neutrino mass matrix (\ref{seesawmass}) and the 
charged-lepton masses introduces in general a mismatch between the mass 
and flavour eigenstates~\cite{MNS}:
\begin{equation}
V_{MNS} \; \equiv \; V_\ell V_\nu^\dagger,
\label{MNS}
\end{equation}
which is reminiscent of the way the CKM matrix appears in the quark 
sector~\cite{EGN}:
\begin{equation}
V_{CKM} \; \equiv \; V_d V_u^\dagger,
\label{CKM}
\end{equation}
though the difference in the ways the quark and neutrino masses 
(\ref{seesaw}) arise may give us some hope that the patterns of neutrino 
and quark mixing, $V_{MNS}$ and $V_{CKM}$, could be somewhat different.

The MNS matrix describing neutrino oscillations can be written in the form
\begin{eqnarray}
V \; = \; \left(
\begin{array}{ccc}
c_{12} & s_{12} & 0 \\
- s_{12} & c_{12} & 0 \\
0 & 0 & 1
\end{array} 
\right)
\left(
\begin{array}{ccc}
1 & 0 & 0 \\
0 & c_{23} & s_{23} \\
0 & - s_{23} & c_{23}
\end{array}
\right)
\left(
\begin{array}{ccc}
c_{13} & 0 & s_{13} \\
0 & 1 & 0 \\
- s_{13} e^{- i \delta} & 0 & c_{13} e^{- i \delta}
\end{array}
\right),
\label{MNSmatrix}
\end{eqnarray}
and there are in addition two CP-violating phases that are not observable 
in neutrino oscillations, but appear in neutrinoless double-$\beta$ decay.

The first matrix factor in (\ref{MNSmatrix}) is measurable in solar
neutrino experiments, and the recent data from SNO~\cite{SNO} and
Super-Kamiokande~\cite{SKsolar} prefer quite strongly the
large-mixing-angle (LMA) solution to the solar neutrino problem with
$\Delta m_{12}^2 \sim 6 \times 10^{-5}$~eV$^2$ and large but non-maximal
mixing: $\theta_{12} \sim 30^o$. The validity or otherwise of the LMA
solution is expected to be settled quite soon by the KamLAND experiment.
The second matrix factor in (\ref{MNSmatrix}) is measurable in atmospheric
neutrino experiments, and the data from Super-Kamiokande in
particular~\cite{SK} favour maximal mixing of atmospheric neutrinos:  
$\theta_{23} \sim 45^o$ and $\Delta m_{23}^2 \sim 2.5 \times
10^{-3}$~eV$^2$. However, the third matrix factor in (\ref{MNSmatrix}) is
basically unknown, with experiments such as Chooz~\cite{Chooz} and
Super-Kamiokande only establishing upper limits on $\theta_{13}$, and {\it
a fortiori} providing no information on the CP-violating phase $\delta$.

\begin{figure}[htb]
\hspace{2cm}
\epsfig{figure=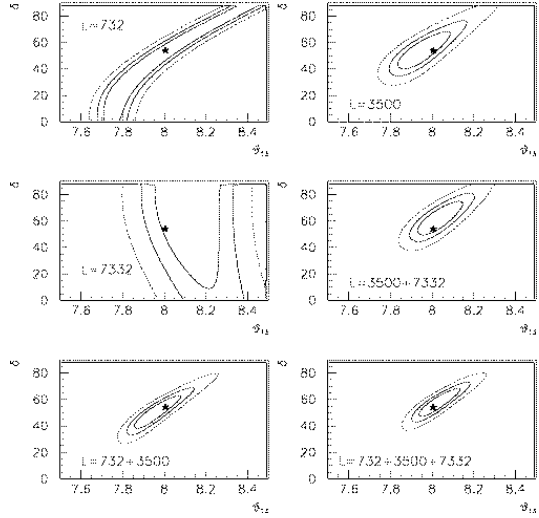,width=7cm}
\hglue3.5cm
\caption[]{\it A simultaneous fit to $\theta_{13}$ and $\delta$,
using a neutrino-factory beam with different baselines and
detector techniques~\cite{golden}, may enable the CP-violating phase
$\delta$ to be extracted.}
\label{fig:cpnu}
\end{figure}

The phase $\delta$ could in principle be measured by comparing the
oscillation probabilities for neutrinos and antineutrinos as seen in
Fig.~\ref{fig:cpnu}~\cite{golden}. This is possible only if $\Delta
m_{12}^2$ and $s_{12}$ are large enough - as now suggested by the success
of the LMA solution to the solar neutrino problem, and if $s_{13}$ is
large enough - which remains an open question.

The effective low-energy mass matrix for the light neutrinos contains 9
parameters, 3 mass eigenvalues, 3 real mixing angles and 3 CP-violating
phases. However, these are not all the parameters in the minimal seesaw
model. In fact, this model has a total of 18 parameters~\cite{Casas,EHLR}.
The remaining 9 associated with the heavy-neutrino sector may be
measurable via their renormalization effects on soft
supersymmetry-breaking parameters, as we discuss below. The total number 
of CP-violating parameters is 6, including the MNS phase $\delta$, the two 
Majorana phases relevant to neutrinoless double-$\beta$ decay, and three 
extra phases that play a key r\^ole in leptogenesis, as we discuss later.

\section{Flavour and CP Violation for Charged Leptons}

\begin{figure}[htb]
\begin{centering}
\epsfig{figure=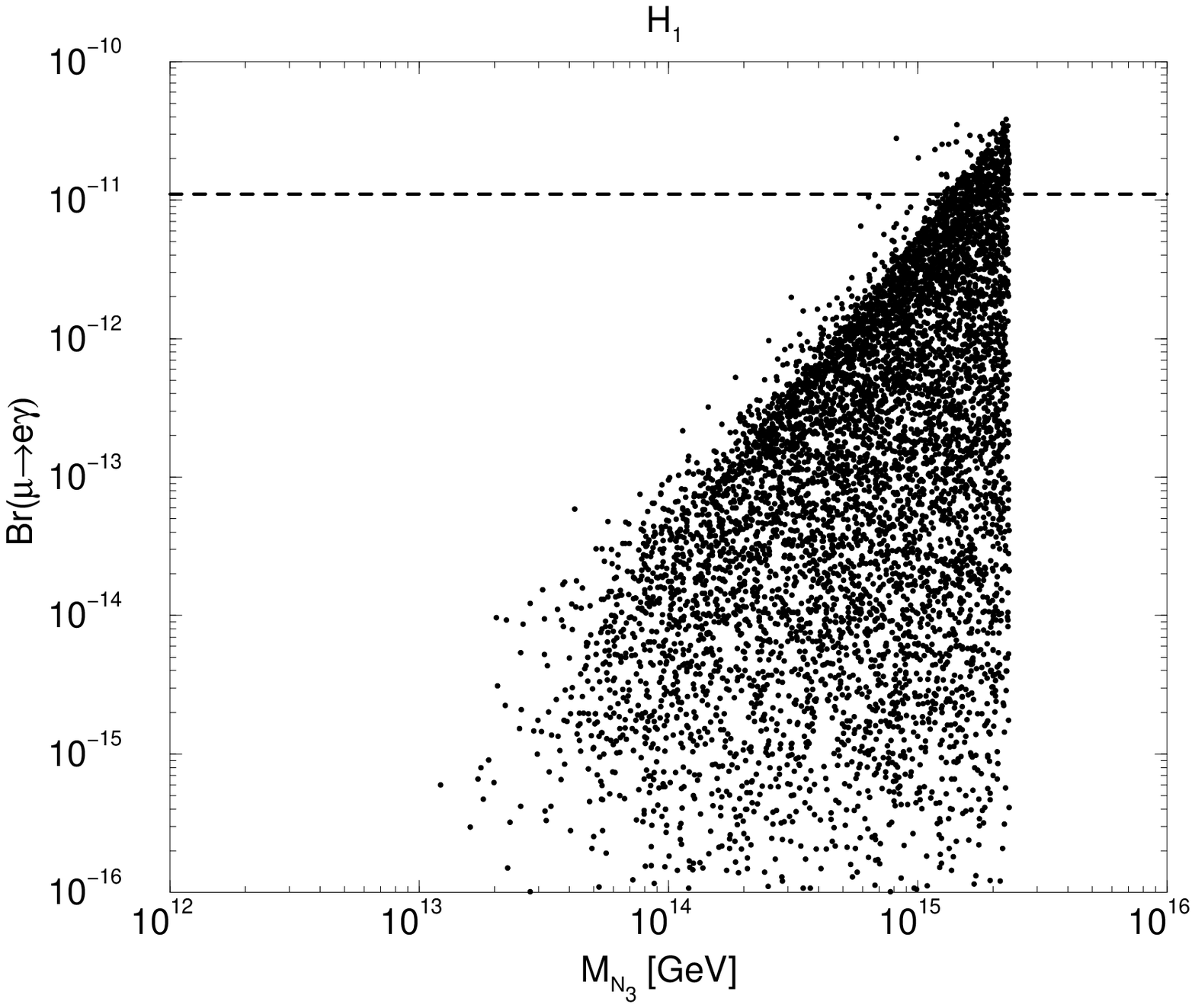,width=4cm}
\hspace{-0.35cm}
\epsfig{figure=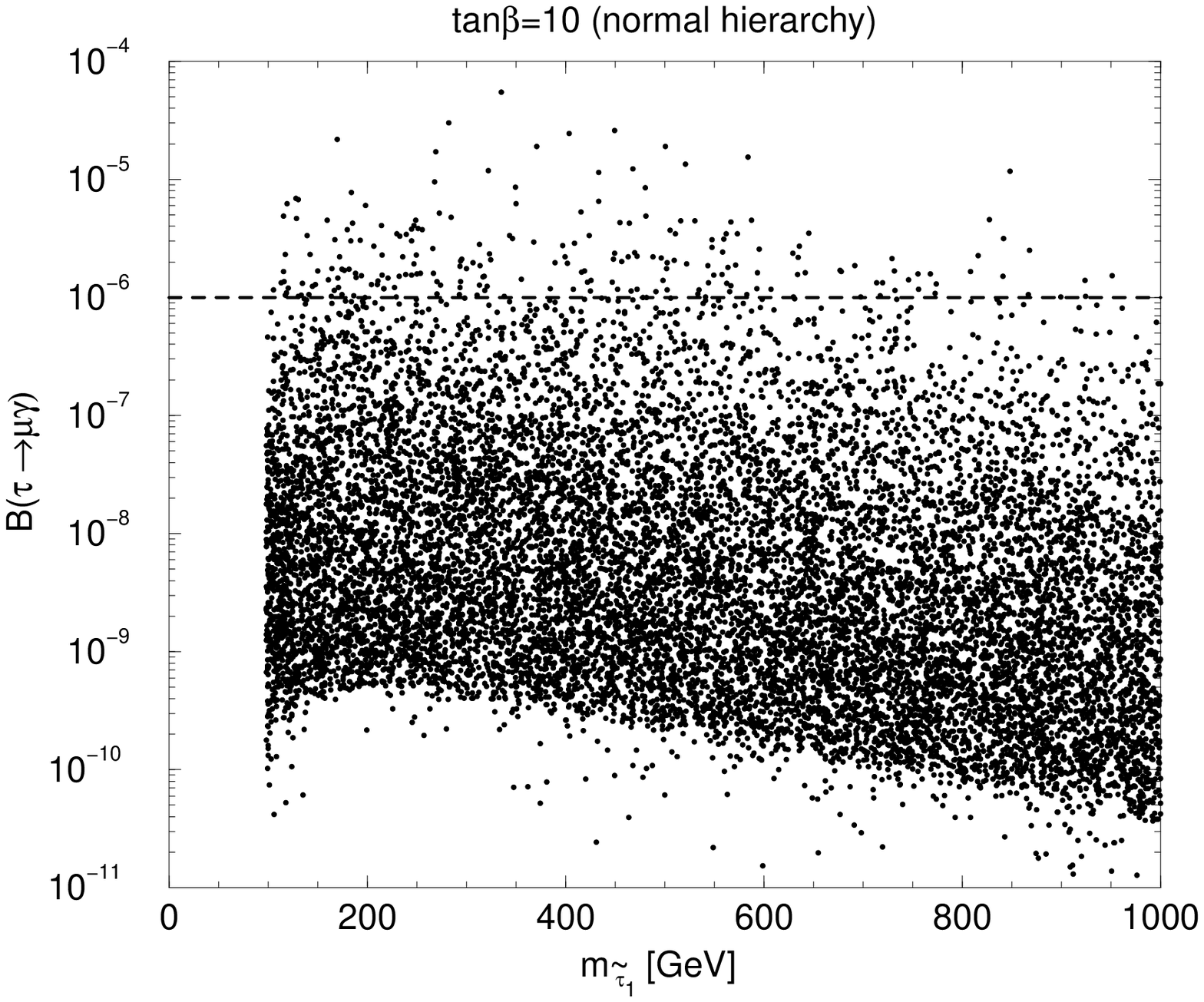,width=4cm}
\hspace{-0.35cm}
\epsfig{figure=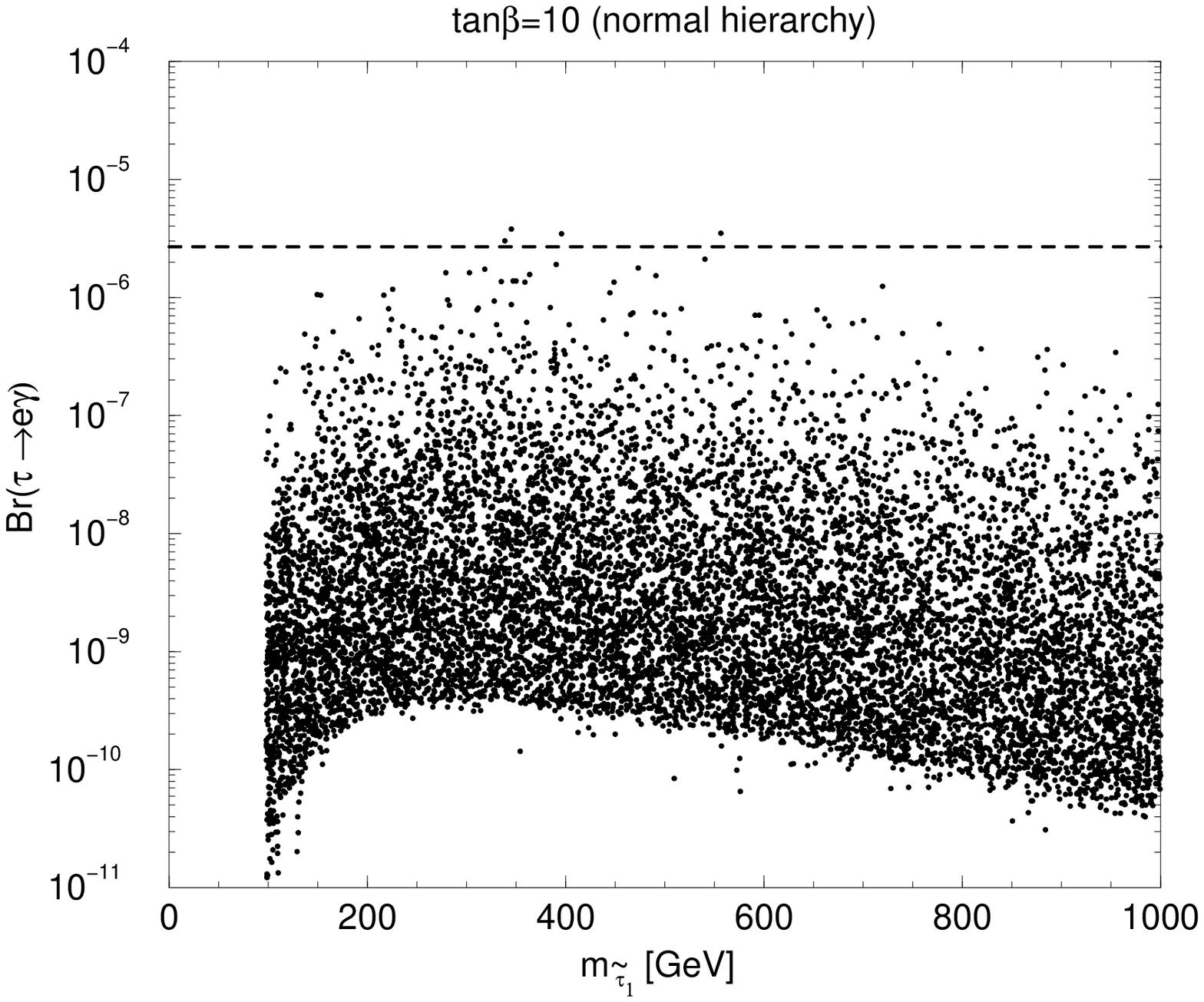,width=4cm}
\end{centering}
\caption[]{\it Scatter plots of the branching ratios for (a) $\mu \to 
e \gamma$, (b) $\tau \to \mu \gamma$ and (c)  $\tau \to e \gamma$ in 
variants of the supersymmetric seesaw model, for various
values of its unknown parameters~\cite{EHRS2}.}
\label{fig:muegamma}
\end{figure}

Assuming that the soft supersymmetry-breaking parameters put it at the 
GUT scale are universal, and working in the leading-logarithmic
approximation with degenerate heavy singlet neutrinos, one finds
the following radiative corrections to the soft
supersymmetry-breaking terms for sleptons:
\begin{eqnarray}
\left( \delta m_{\tilde L}^2 \right)_{ij} \; &=& \;
- { 1 \over 8 \pi^2} \left( 3 m_0^2 + A_0^2 \right) \left( Y_\nu^\dagger
Y_\nu \right)_{ij} {\rm Ln} \left( {M_{GUT} \over M} \right), \nonumber \\
\left( \delta A_\ell \right)_{ij} \; &=& \;
- { 1 \over 8 \pi^2} A_0 Y_{\ell_i} \left( Y_\nu^\dagger
Y_\nu \right)_{ij} {\rm Ln} \left( {M_{GUT} \over M} \right).
\label{leading}
\end{eqnarray}
The non-universality of the corrections (\ref{leading}) leads to processes
that violate the different charged lepton numbers, such as $\mu \to e
\gamma, \tau \to \mu \gamma, \tau \to e \gamma, \mu N \to e N, \mu \to 3 
e, \tau \to 3 e, e 2 \mu, \mu 2 e$ and 3 $\mu$~\cite{EHLR,EHRS1}. 
Fig.~\ref{fig:muegamma}(a) 
shows that the branching ratio for $\mu \to e \gamma$ could be close to the
present experimental upper limit, and Figs.~\ref{fig:muegamma}(b) and (c) 
makes the same point for the decays $\tau \to \mu 
\gamma$ and $\tau \to e \gamma$, respectively~\cite{EHRS2}.

\begin{figure}[htb]
\hspace{1cm}
\epsfig{figure=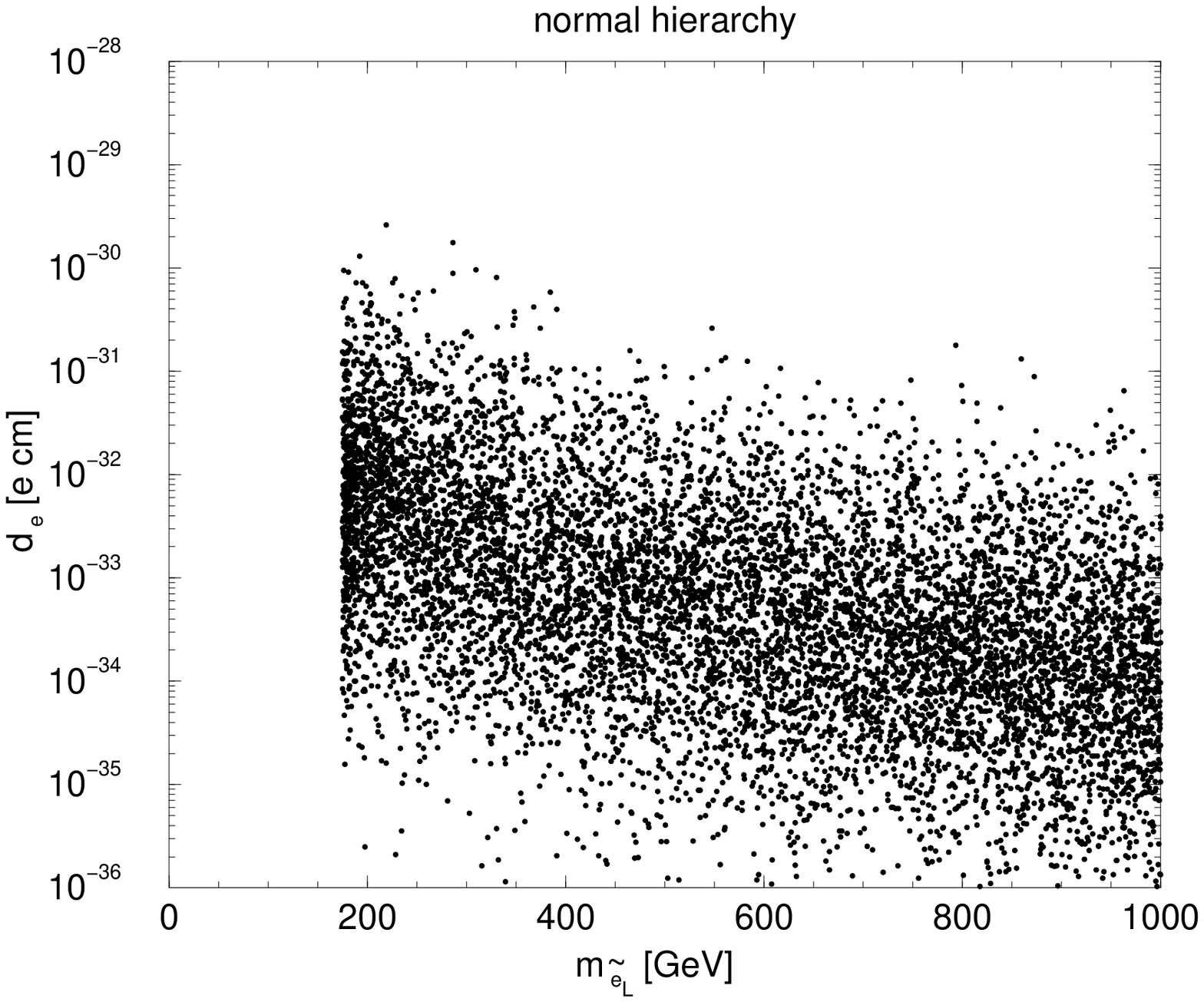,width=5cm}
\epsfig{figure=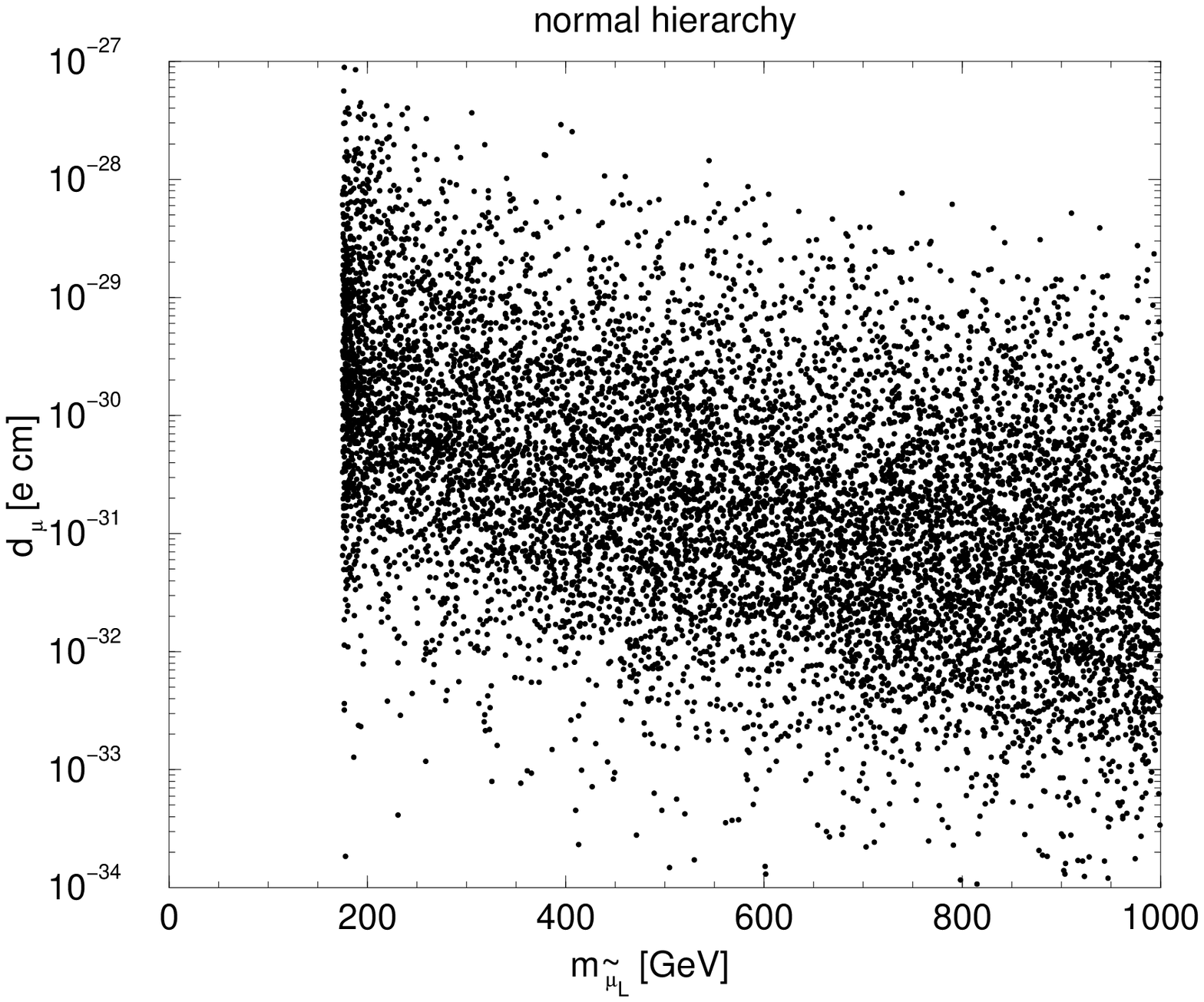,width=5cm}
\hglue3.5cm   
\caption[]{\it Scatter plots of (a) $d_e$ and (b) $d_\mu$ in variants of
the supersymmetric seesaw model, for different values of the
unknown parameters~\cite{EHRS2}.}
\label{fig:dedmu}
\end{figure}

The electric dipole moments of the electron and muon depend sensitively on
the non-degeneracy of the heavy singlet neutrinos~\cite{EHRS1,EHRS2}. As
seen in Fig.~\ref{fig:dedmu}, they could take values as large as $d_e \sim
3 \times 10^{-30}$~e.cm and $d_\mu \sim 10^{-27}$~e.cm, to be compared
with the present experimental upper limits of $d_e < 1.6 \times
10^{-27}$~e.cm~\cite{de} and $d_\mu < 10^{-18}$~e.cm~\cite{BNL1}. An
ongoing series of experiments might be able to reach $d_e \sim 3 \times
10^{-30}$~e.cm, and a type of solid-state experiment that might be
sensitive to $d_e \sim 10^{-33}$~e.cm has been proposed~\cite{Lamoreaux}.
Also, $d_\mu \sim 10^{-24}$~e.cm might be accessible with the PRISM
experiment proposed for the JHF~\cite{PRISM}, and $d_\mu \sim 5 \times
10^{-26}$~e.cm might be attainable at the front end of a neutrino
factory~\cite{nufact}.

\section{Leptogenesis}

One of the favoured scenarios for baryogenesis is first to generate a 
lepton asymmetry via CP-violating decays of heavy singlet neutrinos, which 
is then recycled into a baryon asymmetry via non-perturbative electroweak 
interactions~\cite{FY}. The CP asymmetry in this leptogenesis scenario is 
related to the product $Y_\nu Y^\dagger_\nu$.
The total decay rate of a heavy neutrino $N_i$ may be written in the
form
\begin{equation}
\Gamma_i \; = \; {1 \over 8 \pi} \left( Y_\nu Y^\dagger_\nu \right)_{ii}
M_i,
\label{gammai}
\end{equation}
and one-loop CP-violating diagrams involving the exchange of heavy
neutrino $N_j$ would generate an asymmetry in $N_i$ decay of the form:  
\begin{equation}
\epsilon_{ij} \; = \; {1 \over 8 \pi} {1 \over \left( Y_\nu Y^\dagger_\nu
\right)_{ii}} {\rm Im} \left( \left( Y_\nu Y^\dagger_\nu \right)_{ij}
\right)^2 f \left( {M_j \over M_i} \right),
\label{epsilon}
\end{equation}
where $f ( M_j / M_i )$ is a known kinematic function.

\begin{figure}[htb]
\begin{centering}
\hspace{2cm}
\epsfig{figure=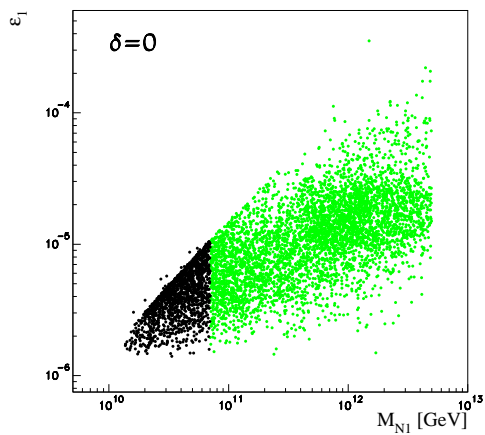,width=7cm}
\end{centering}
\hglue3.5cm
\caption[]{\it Heavy singlet neutrino decay may exhibit a CP-violating
asymmetry, leading to leptogenesis and hence baryogenesis, even if the
neutrino oscillation phase $\delta$ vanishes~\cite{ER}.}
\label{fig:nodelta}
\end{figure}

The relevant combination $Y_\nu Y^\dagger_\nu$ is independent of $V_{MNS}$
and hence of the light neutrino mixing angles and CP-violating phases. The
basic reason for this is that one makes a unitary sum over all the light
lepton species in evaluating the decay asymmetry $\epsilon_{ij}$
(\ref{epsilon}). Fig.~\ref{fig:nodelta} shows explicitly that one can 
generate a lepton asymmetry even if the MNS phase $\delta$ vanishes.

In general, one may formulate the following strategy for calculating
leptogenesis in terms of laboratory observables:

$\bullet$ Measure the neutrino oscillation phase $\delta$ and the Majorana   
phases,

$\bullet$ Measure observables related to the renormalization of soft
supersymmetry-breaking parameters, that are functions of $\delta$,
the Majorana and leptogenesis phases,

$\bullet$ Extract the effects of the known values of $\delta$ and
the Majorana phases, and thereby isolate the leptogenesis parameters.

\section{CP Violation in the MSSM Higgs Sector}

A popular alternative scenario for baryogenesis has been to generate a
quark asymmetry at the electroweak scale~\cite{FS}. This requires a
breakdown of thermal equilibrium, necessitating a first-order electroweak
phase transition. This is impossible in the SM, since LEP tells us that
the Higgs boson weighs more than $114.4$~GeV, whereas a first-torder
electroweak phase transition is possible only if $m_H <
70$~GeV~\cite{Laine}. Generating a first-order phase transition would
require extra light scalar bosons, as could be provided in supersymmetry,
if the lighter $\tilde t$ is very light. This scenario would also require
more CP violation than is present in the SM.

Indeed, two extra CP-violating phases appear in the MSSM, even if the soft 
supersymmetry-breaking parameters are universal at the input GUT scale, as 
assumed here. These can be taken as the (supposedly common) phases of the 
trilinear soft supersymmetry-breaking parameters Arg($A_{t,b}$) and the 
phase of the 
gluino mass Arg($m_{\tilde g}$). These generate mixing between the 
`scalar'and `pseudoscalar' MSSM Higgs bosons: at the one-loop level
\begin{equation}
\delta m^2_{SP} \; \sim \; {m_t^4 \over v^2} {\mu {\rm Im} A_t \over 32 
\pi^2 m^2_{susy}} + \cdots,
\label{CPmixing}
\end{equation}
and a dependence on Arg($m_{\tilde g}$) appears at the two-loop level.

\begin{figure}
\hspace{1cm}
\epsfig{figure=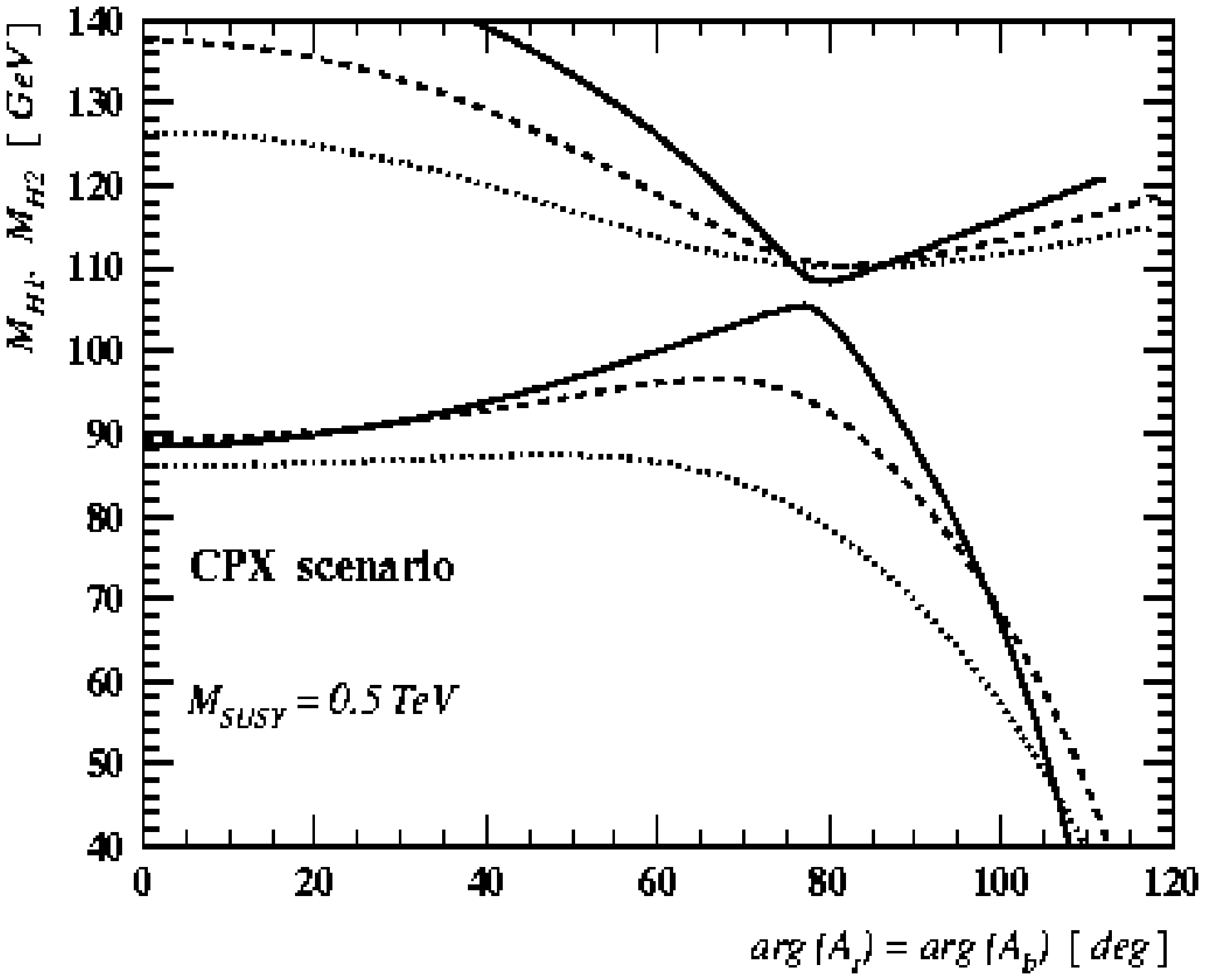,width=5cm}
\epsfig{figure=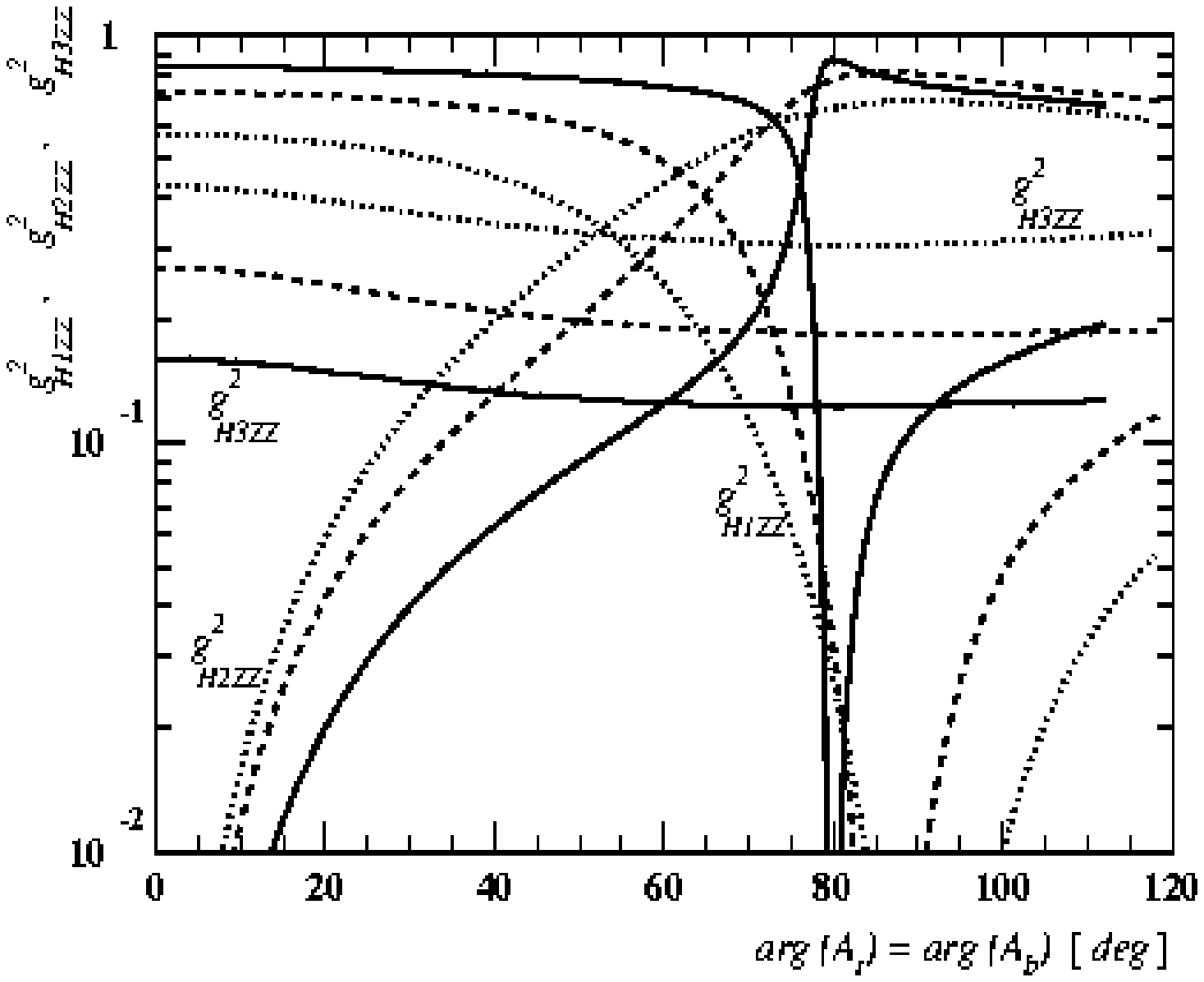,width=5cm}
\hglue3.5cm   
\caption[]{\it In the MSSM with maximal CP violation in the Higgs sector, 
(a) there may be level-crossing between the lightest and second-lightest 
Higgs bosons, and (b) the lightest Higgs boson may have a small coupling 
in the process $e^+ e^- \to Z + H$~\cite{CEPW}.} 
\label{fig:CPHiggs}
\end{figure}

In the presence of CP violation, it is convenient to parametrize the MSSM
Higgs sector in terms of $m_{H^+}$ and $\tan \beta$. As seen in
Fig.~\ref{fig:CPHiggs}~\cite{CEPW}, there may be level crossing between
the two lightest neutral Higgs bosons, and the lightest Higgs $H_1$ may
have a suppressed coupling in the process $e^+ e^- \to Z + H_1$. In this
case, it could be that there exists a light Higgs boson lurking below the
lower limit established by LEP in the SM. The prospects that experiments
at hadron colliders may be able to plug this hole are discussed
in~\cite{CEMPW}.

The phenomenology of CP-violating Higgs bosons is very rich, and only its
surface has been scratched. A $\mu^+ \mu^-$ collider - either at the
energy of the lightest Higgs boson, or close to the nearby masses of the
second and third neutral Higgs bosons - may be necessary one day to
unravel this physics~\cite{mumu}.

\section{Some Answers}

At a round-table discussion earlier this week, some central
questions were raised, to which I would like to provide some 
personal answers.
\begin{itemize}
\item{{\it Q: What is the r\^ole of flavour studies in providing clues 
about 
new physics?} \\
A: They may cast light on the darkest corners of supersymmetry, 
namely its flavour and CP problems.}
\item{{\it Q: What are the implications of CP studies for our 
understanding of baryogenesis?} \\
A: Standard Model CP violation is inadequate for the task, 
but CP violation in either the lepton or Higgs sector could do the job. 
Both may be tested in future experiments.}
\item{{\it Q: What are the implications of lepton mixing for unification 
and phenomenology?} \\
A: It provides a direct window on physics at the GUT scale, 
and could open up a whole new arena for experiments on decays that violate 
the charged lepton flavours, such as $\mu \to e \gamma, \tau \to \mu 
\gamma, \tau \to e \gamma$ and many more.}
\end{itemize}
Flavour physics and CP violation surely have a long and glorious future!

\end{document}